\documentclass[conference]{IEEEtran}
\IEEEoverridecommandlockouts
\usepackage{cite}
\usepackage{amsmath,amssymb,amsfonts}
\usepackage{algorithmic}
\usepackage{graphicx}
\usepackage{textcomp}
\usepackage{xcolor}
\usepackage{booktabs}
\usepackage{array}
\usepackage{stfloats}
\usepackage{tikz}
\usetikzlibrary{arrows.meta,positioning,fit,backgrounds,calc}
\def\BibTeX{{\rm B\kern-.05em{\sc i\kern-.025em b}\kern-.08em
    T\kern-.1667em\lower.7ex\hbox{E}\kern-.125emX}}

\usepackage{eso-pic}
\AddToShipoutPictureBG*{%
  \AtPageLowerLeft{%
    \put(0,25){\makebox[\paperwidth]{\parbox{0.85\paperwidth}{\scriptsize\centering
    \copyright~2026 IEEE. Personal use of this material is permitted. Permission from IEEE must be obtained for all other uses, in any current or future media, including reprinting/republishing this material for advertising or promotional purposes, creating new collective works, for resale or redistribution to servers or lists, or reuse of any copyrighted component of this work in other works.}}}}}

\begin{document}

\title{Characterizing Frozen SSL Encoders and Backends for
Cross-Domain Speech Deepfake Detection}

\author{\IEEEauthorblockN{Benjamin Hurt\IEEEauthorrefmark{1},
Oscar O'Donnell\IEEEauthorrefmark{2}\IEEEauthorrefmark{3},
Yefei Yang\IEEEauthorrefmark{1}\IEEEauthorrefmark{3},
Anjani Sethi\IEEEauthorrefmark{1},
Priyanka Durai\IEEEauthorrefmark{1}}
\IEEEauthorblockA{\IEEEauthorrefmark{1}University of California, Santa Barbara, USA \\
\IEEEauthorrefmark{2}Columbia University, New York, USA \\
\IEEEauthorrefmark{3}These authors contributed equally to this work. \\
Email: \{hurt, yefei\_yang, anjanisethi, priyankadurai\}@ucsb.edu, ojo2013@columbia.edu}}

\maketitle

\begin{abstract}
Self-supervised learning (SSL) encoders are the dominant front end for speech
deepfake detection, and the prevailing recipe fine-tunes the full encoder on a
small spoofing corpus, maximizing in-domain accuracy at the cost of known
degradation under domain shift. Taking a recent cross-attention fusion of SSL and
constant-Q cepstral features as a fixed reference recipe, we freeze its encoder
and ask which encoder and which backend behave best across domains. We cross two
frozen encoders, the massively multilingual XEUS and Wav2Vec2 XLSR-53, with two
backends, multi-head factorized attentive (MHFA) pooling and AASIST, and evaluate
on ASVspoof 2019 LA, ASVspoof 2021 LA and DF, In-the-Wild (ITW), and ASVspoof 5,
training every system in one pipeline over three seeds. We report three
findings. First, encoder and backend interact: MHFA suits the stronger encoder,
while AASIST recovers more from and stabilizes the weaker one. Second, within
this recipe, the frozen encoder produces large, repeatable differences. With
MHFA, frozen XEUS lowers error relative to frozen XLSR-53 by 58--77\% on LA21,
DF21, and ASV5, and more modestly on ITW. This ordering persists in a
matched-aggregation ablation, and a capacity-matched WavLM Large diagnostic
shows that parameter count alone does not explain XLSR-53's weaker performance.
Third, against a three-seed fine-tuned baseline, frozen XEUS updates under 0.4M
parameters versus 316M, wins on ASVspoof 5 (in EER and min-DCF) and modestly on
LA21, trails on DF21 and ITW, and shows lower seed variation on three of the
four out-of-domain sets.
\end{abstract}


\section{Introduction}
Advances in text-to-speech and voice conversion have made synthetic speech
difficult to distinguish from genuine recordings, motivating a large body of
work on speech deepfake detection and anti-spoofing. The field has converged on
a standard architecture in which a large self-supervised learning (SSL) speech
encoder produces frame-level representations that a lightweight backend
classifies as bona fide or spoofed \cite{tak,wangyama}. The dominant recipe
fine-tunes the SSL encoder end to end on a spoofing corpus, typically ASVspoof
2019 LA. Fine-tuning yields excellent in-domain equal error rates (EERs), but it
adapts a general-purpose encoder to the narrow statistics of a single corpus,
and detectors trained this way are widely reported to overfit to known attacks
and to transfer poorly to unseen attacks, codecs, and recording conditions
\cite{tak,liu2023asvspoof21,muller2022}. The In-the-Wild benchmark makes the gap
concrete: detectors that perform well on ASVspoof degrade by up to an order of
magnitude on found real-world audio \cite{muller2022}.

A recent fusion framework, which we refer to as the cross-attention baseline
following its originators \cite{tvot}, combines a fine-tuned Wav2Vec2 XLSR-53
front end with handcrafted constant-Q cepstral coefficients (CQCC) through
cross-attention and an AASIST backend. Its authors identify two directions for
future work: exploring ``different backend classifiers'' and adding
\emph{prosodic} descriptors to the fusion \cite{tvot}. We take up the first
directly and use the second as motivation. Rather than bolting prosodic features
onto the existing encoder, we ask whether an encoder whose pretraining already
spans prosodically diverse material, XEUS \cite{xeus} (over one million hours
across 4{,}057 languages, with singing voice in the corpus and a dereverberation
objective for acoustic robustness), is a better fixed representation than
XLSR-53 \cite{xlsr} when both are frozen. A recent study on AI-generated singing
found XEUS outperformed Wav2Vec among speech embeddings, and its authors
attributed this to a pretraining distribution capturing prosody and timbre
\cite{doubleentendre}.

Freezing the encoder reframes the problem from \emph{adaptation} to
\emph{characterization}. With the encoder fixed, differences across systems
reflect (i) what a given pretraining distribution captures and (ii) how a
backend reads it. A non-fine-tuned front end has been shown to remain
competitive in some settings \cite{wangyama}, yet frozen SSL detectors still
struggle under domain shift, with accuracy varying markedly by encoder and test
corpus \cite{serrano2025}. This is what a shared-pipeline comparison of frozen
encoders can characterize, and it responds to the recognition that a low error
rate alone does not establish trustworthiness \cite{afchar2024,wedefense}. We
freeze rather than adapt with LoRA or adapters \cite{hu2022lora} because any
adaptation re-entangles pretraining with corpus-specific tuning. A frozen encoder compares the representations as they were pretrained. Freezing removes
corpus-specific adaptation from the comparison, but it does not separate
pretraining distribution from objective or scale. Section~\ref{sec:itw} probes
scale with a capacity-matched encoder. Our frozen system accordingly does not
target state-of-the-art EER; it asks which representations transfer across
datasets and with how many trainable parameters, and our claims are scoped
accordingly.

Our contributions are:
\begin{itemize}
\item A shared-pipeline characterization that holds a published cross-attention
fusion recipe \cite{tvot} fixed and crosses two frozen SSL encoders (XEUS,
XLSR-53) with two backends (MHFA, AASIST), across five benchmarks spanning
in-domain, near-domain, real-world, and large-scale crowdsourced conditions.
\item Evidence that encoder and backend interact rather than contribute
independently: the stronger backend depends on the encoder it reads (MHFA suits
the stronger encoder, while AASIST recovers more from and stabilizes the
weaker), so backend choice does not transfer across encoders.
\item Evidence that, within this fixed recipe, frozen encoder choice produces
large and repeatable cross-dataset differences whose ordering persists in a
matched-aggregation ablation. A capacity-matched WavLM Large diagnostic further
shows that XLSR-53's weaker performance is not explained by parameter count
alone. We state pretraining/evaluation provenance overlap explicitly as a limit
on reading these differences as pure distributional generalization.
\item An in-pipeline comparison against a three-seed full-encoder fine-tuned
baseline: the frozen broad-coverage encoder updates under 0.4M parameters versus
316M, performs better on LA21 and ASV5 but worse on DF21 and ITW, and has lower
observed seed variation on three of four out-of-domain sets.
\end{itemize}

\section{Related Work}
\textbf{SSL front ends for anti-spoofing.} Wav2Vec2 XLSR-53 \cite{xlsr} and
related models, typically fine-tuned, underpin most recent detectors; the
wav2vec2-plus-AASIST recipe of Tak et al.\ \cite{tak} is representative and was
motivated by the poor generalization of conventional countermeasures. Wang and
Yamagishi \cite{wangyama} report that a non-fine-tuned front end can be
competitive while still favoring fine-tuning for best in-domain results,
motivating our frozen design. XEUS \cite{xeus} is a more recent massively
multilingual encoder whose pretraining emphasizes acoustic robustness. To our
knowledge, its only prior anti-spoofing appearance is as a front end reported
not to improve results, with no published number, in a front-end comparison
\cite{aasist3}, leaving a systematic frozen comparison against XLSR-53 open. A
recent benchmark of twenty frozen SSL front ends finds large multilingual
discriminative encoders among the strongest and adopts the frozen-encoder,
simple-backend paradigm we use, but omits XEUS \cite{spoofsuperb}.

\textbf{Backends.} AASIST \cite{aasist} couples spectro-temporal graph attention
with attentive statistics pooling and is the backend in the cross-attention
baseline \cite{tvot}. MHFA \cite{mhfa}, introduced for speaker verification,
learns separate key and value weightings over encoder layers and pools with
multiple factorized heads. It was proposed as an efficient backend over
pretrained encoders. We compare the two inside one fusion architecture.

\textbf{Layer aggregation.} SSL layers encode different information
\cite{pasad2021}, and for deepfake detection the most discriminative spoofing
cues concentrate in lower-to-middle transformer layers \cite{elkheir2025layer}.
This motivates treating layer aggregation as part of each encoder's
configuration rather than as a free variable, and we isolate its effect in
Section~\ref{sec:ablation}.

\textbf{Parameter-efficient detection.} Compact backends over frozen foundation
models are an active direction. Nes2Net \cite{nes2net} argues that
dimensionality-reduction layers in backends lose information at added parameter
cost, and proposes a lighter nested design. Nes2Net benchmarks backends over a fixed encoder, whereas our axis is the orthogonal one: which frozen encoder to fix,
studied factorially with a matched-aggregation ablation.

\textbf{Domain generalization.} A complementary line of work attacks domain
shift through explicit feature alignment: corpus-invariant features via gradient
reversal \cite{dao2026}, unsupervised statistical adaptation of SSL embeddings
\cite{thani2026}, and the multilingual ML-ITW benchmark documenting degradation
under real-world conditions \cite{li2026mlitw}. Whether out-of-domain failure
reflects genuinely harder attacks or merely different ones has itself been
studied directly \cite{harderdiff}. These post-hoc remedies yield modest, uneven gains. We instead ask whether the pretraining distribution can be chosen so that
robustness is a property of the frozen representation rather than of a
corrective adaptation step.

\section{Method}

\begin{figure}[t]
\centering
\begin{tikzpicture}[
  font=\scriptsize,
  >={Stealth[length=2mm]},
  box/.style={draw, rounded corners=2pt, align=center, inner sep=3pt,
              minimum height=6.5mm, text width=26mm},
  wide/.style={draw, rounded corners=2pt, align=center, inner sep=3pt,
               minimum height=6.5mm, text width=66mm},
  mid/.style={draw, rounded corners=2pt, align=center, inner sep=3pt,
              minimum height=6.5mm, text width=40mm},
  frozen/.style={fill=blue!8},
  trained/.style={fill=orange!12},
  io/.style={fill=black!5},
  arr/.style={->, thick, black!70}
]
\node[wide, io]      (wav)  at (0,0)      {Input waveform};
\node[box, frozen]   (enc)  at (-2.1,-1.2){Frozen SSL encoder\\(XEUS \emph{/} XLSR-53)};
\node[box, io]       (cqcc) at (2.1,-1.2) {CQCC (60-dim)};
\node[box, trained]  (pssl) at (-2.1,-2.4){Proj.\ $1024{\to}128$};
\node[box, trained]  (psf)  at (2.1,-2.4) {Proj.\ $60{\to}128$};
\node[wide, trained] (ca)   at (0,-3.6)   {Cross-attention fusion (Eq.~\eqref{eq:ca})};
\node[mid, trained]  (back) at (0,-4.7)   {Backend: MHFA \emph{/} AASIST};
\node[mid, trained]  (head) at (0,-5.7)   {LayerNorm $+$ Linear};
\node[box, io]       (out)  at (0,-6.8)   {bona fide / spoof};
\draw[arr] (wav.south) -- ++(0,-0.3) -| (enc.north);
\draw[arr] (wav.south) -- ++(0,-0.3) -| (cqcc.north);
\draw[arr] (enc)  -- (pssl);
\draw[arr] (cqcc) -- (psf);
\draw[arr] (pssl.south) -- (pssl.south |- ca.north);
\draw[arr] (psf.south)  -- (psf.south  |- ca.north);
\draw[arr] (ca)   -- (back);
\draw[arr] (back) -- (head);
\draw[arr] (head) -- (out);
\node[draw, fill=blue!8, minimum width=3mm, minimum height=3mm,
      inner sep=0pt, label=right:{frozen}] at (-1.6,-7.6) {};
\node[draw, fill=orange!12, minimum width=3mm, minimum height=3mm,
      inner sep=0pt, label=right:{trained}] at (0.7,-7.6) {};
\end{tikzpicture}
\caption{Frozen-encoder cross-attention fusion. The SSL encoder (blue) is
frozen; only the projections, fusion, backend, and classifier (orange) are
trained. The SSL stream queries the spectral stream and a residual preserves
the SSL summary (Eq.~\eqref{eq:ca}). We cross two encoders with two backends,
giving the four systems studied.}
\label{fig:arch}
\end{figure}
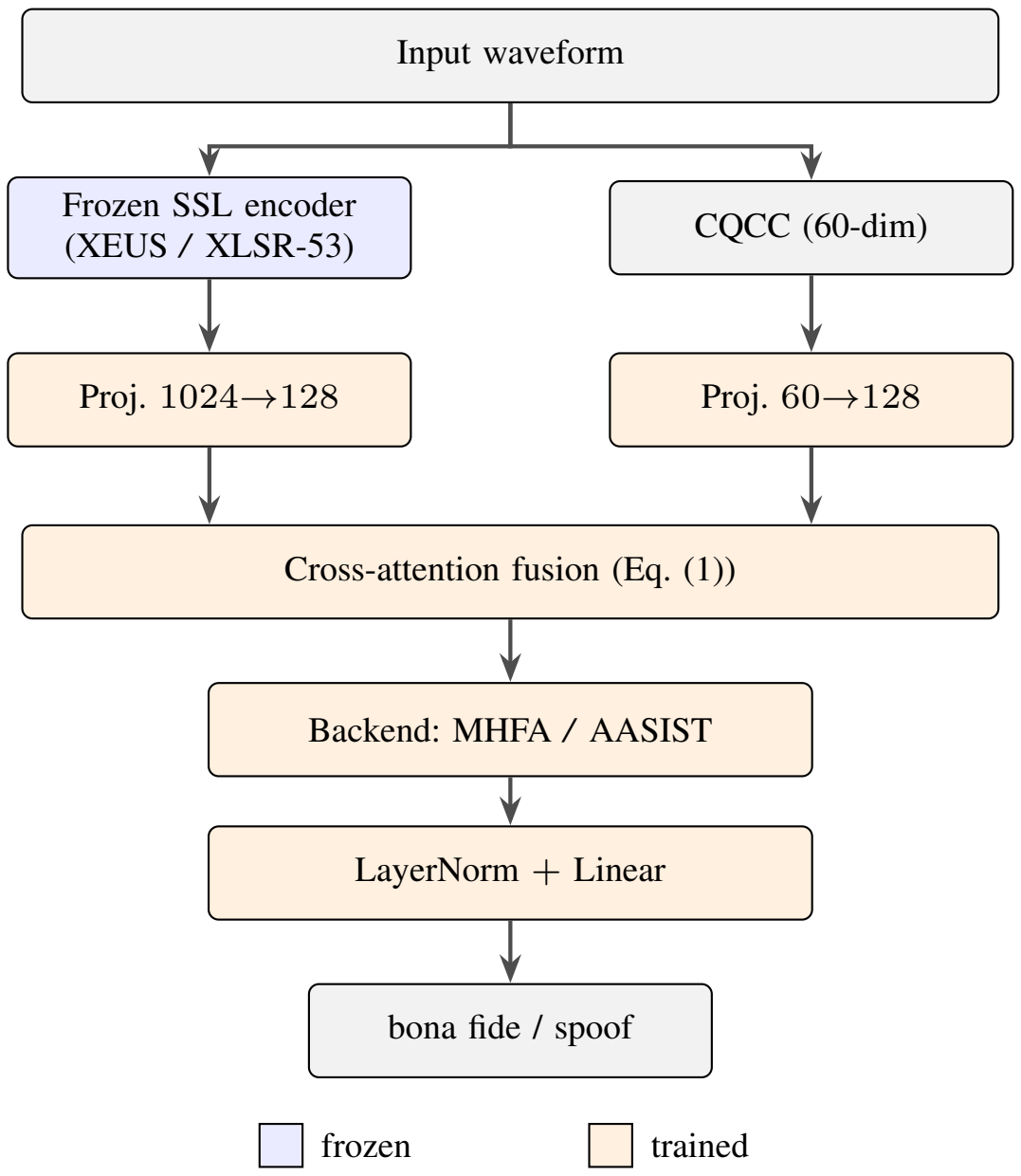

\subsection{Cross-attention fusion architecture}
The architecture is shown in Fig.~\ref{fig:arch}. For an input waveform we
extract two streams. The SSL stream applies a frozen encoder
$E\in\{\text{XEUS},\text{XLSR-53}\}$ to obtain frame features
$\mathbf{f}_{\text{ssl}}\in\mathbb{R}^{T_{\text{ssl}}\times 1024}$ (1024 for
both encoders). The spectral stream computes 60-dimensional CQCC features
\cite{cqcc} (20 cepstral coefficients from a constant-Q transform at 96 bins per
octave, with $\Delta$ and $\Delta\Delta$),
$\mathbf{f}_{\text{sf}}\in\mathbb{R}^{T_{\text{sf}} \times 60}$. Linear
projections map both streams to a common dimension $D{=}128$, and the spectral
stream is length-matched to the SSL frame count $T_{\text{ssl}}{=}T$ by linear
interpolation, giving
$\tilde{\mathbf{f}}_{\text{ssl}},\tilde{\mathbf{f}}_{\text{sf}}
\in\mathbb{R}^{T\times D}$.

Fusion is cross-attention with the projected SSL stream as query and the
projected spectral stream as key and value, with a residual connection
preserving the SSL summary:
\begin{equation}
\mathbf{H} = \mathrm{Softmax}\!\left(\frac{\mathbf{Q}
\mathbf{K}^{\top}}{\sqrt{D}}\right)\mathbf{V}
+ \tilde{\mathbf{f}}_{\text{ssl}},
\label{eq:ca}
\end{equation}
where $\mathbf{Q}{=}\tilde{\mathbf{f}}_{\text{ssl}}W_Q$,
$\mathbf{K}{=}\tilde{\mathbf{f}}_{\text{sf}}W_K$, and
$\mathbf{V}{=}\tilde{\mathbf{f}}_{\text{sf}}W_V$ with
$W_Q,W_K,W_V\in\mathbb{R}^{D\times D}$, so every term is
$\mathbb{R}^{T\times D}$ and the residual is dimensionally consistent. The fused
sequence $\mathbf{H}$ is passed to a backend that produces an utterance
embedding, followed by layer normalization and a linear bona-fide/spoof
classifier.

\subsection{Backends}
We compare two backends operating on $\mathbf{H}$. \emph{MHFA} \cite{mhfa} forms
keys and values, pools with $H{=}8$ factorized heads over $K{=}64$ compressed
components, and emits a 128-dimensional embedding. \emph{AASIST} \cite{aasist}
applies its spectro-temporal graph attention stack. Both are lightweight
relative to the encoder.

\subsection{Frozen factorial design}
\label{subsec:design}
We cross encoder $\in\{$XEUS, XLSR-53$\}$ with backend $\in\{$MHFA, AASIST$\}$,
giving four systems. In every cell the encoder is frozen and only the
projections, cross-attention, backend, and classifier are trained. We are
deliberate about what is and is not matched. Matched across cells: the encoder's
frozen status, the CQCC spectral branch, preprocessing, the fusion mechanism of
Eq.~\eqref{eq:ca}, and the evaluation protocol; encoder and backend are the two
experimental factors. Not matched, and treated as part of each encoder's
best-practice configuration rather than as controlled variables: the SSL
layer-aggregation (XEUS final layer versus XLSR-53 mean of the last four, chosen
per encoder following layer-wise evidence \cite{pasad2021,elkheir2025layer}),
training epochs, and the MHFA-cell learning-rate schedule. We therefore present
the main factorial as a comparison of each encoder in a reasonable
configuration, not as a single-variable ablation. Section~\ref{sec:ablation} isolates the layer-aggregation factor, and the remaining factors are discussed
in Section~\ref{sec:limits}. We hold the cross-attention CQCC fusion fixed
because it is the baseline's empirically best variant \cite{tvot}; ablating the
spectral branch is orthogonal to the encoder question.

We use XLSR-53 as the comparator because it is the exact front end fixed by the
reference recipe \cite{tvot}. A consequence we make explicit: our encoder
comparison is internal to this recipe. It speaks to what swapping the encoder
does within a fixed published system, not to which frozen encoder is best in
general, a question we return to in Section~\ref{sec:itw}.

\section{Experimental Setup}
\subsection{Data and protocol}
All systems are trained on ASVspoof 2019 LA (LA19) \cite{asvspoof2019} with no
data augmentation, matching the baseline's no-augmentation setting. We evaluate
on five benchmarks of increasing distance from training: LA19 (in domain);
ASVspoof 2021 LA (LA21; near domain, codec and transmission effects); ASVspoof
2021 DF (DF21; compressed deepfake audio) \cite{liu2023asvspoof21}; In-the-Wild
(ITW; 37.9 hours of found celebrity and politician recordings)
\cite{muller2022}; and ASVspoof 5 (ASV5; large-scale crowdsourced data with 16
eval attacks) \cite{wang2024asvspoof5}. We use the same preprocessing as the
baseline \cite{tvot}: 16\,kHz resampling, pre-emphasis (0.97), fixed-length
cropping/padding to 64{,}600 samples, and 25\,ms\,/\,10\,ms framing. Keeping the
front end identical means differences across systems reflect the encoder and
backend rather than signal conditioning. Every system, including the fine-tuned
baseline, is trained and scored in one pipeline over three seeds.

\subsection{Metrics}
We report pooled EER (\%) as mean$\pm$standard deviation over three seeds
(\{42, 1234, 2025\}). Per-seed EERs use a single tie-robust convention, and 95\%
confidence intervals (class-stratified bootstrap percentile intervals over
evaluation utterances) quantify within-run scoring uncertainty where cited. EER
is threshold-free and directly comparable across all five benchmarks. On ASV5,
the one set with a standardized cost function, we additionally report the
official countermeasure min-DCF ($P_{\text{spoof}}{=}0.05$,
$C_{\text{miss}}{=}1$, $C_{\text{fa}}{=}10$). a-DCF requires
speaker-verification scores that lie outside the scope of this countermeasure
study.

\subsection{Baseline and parameter accounting}
To compare frozen detection against full fine-tuning under the same data and
evaluation pipeline, we train an in-pipeline fine-tuned baseline: XLSR-53 (315M
parameters\footnote{315M for the bare encoder; ${\sim}$316M including the AASIST
head.}) with all encoder weights unfrozen plus an AASIST backend, trained on
LA19 and scored over the same three seeds and convention as the frozen systems.
Each fine-tuned seed was selected by best development EER (0.12\% for seed 42),
an out-of-domain-blind rule (Section~\ref{sec:efficiency}). Our frozen fusion
heads have 390k trainable parameters (MHFA cells) and 332k (AASIST cells),
versus the ${\sim}$316M updated by full fine-tuning, a reduction of nearly three
orders of magnitude (about 800-fold). We are explicit that this is a
\emph{trainable-parameter} statement, not a model-size or measured-runtime
statement: XEUS is itself a large encoder (577M parameters), and the inference
footprint is comparable to, or larger than, a fine-tuned XLSR-53 system. The
claim concerns what must be optimized, not what must be stored or run, in line
with the ${\sim}10^5$-parameter trainable heads reported over frozen models
\cite{serrano2025}.

\section{Results and Analysis}

\begin{table*}[t]
\caption{Pooled EER (\%), mean$\pm$std over three seeds (\{42, 1234, 2025\}), by
frozen encoder and backend, all trained and scored in one pipeline (lower
better). The fine-tuned baseline (XLSR-53 unfrozen + AASIST) is also three-seed
but not frozen. \textbf{Bold}/$\dagger$ refer to the four factorial systems:
bold marks the best per column, $\dagger$ the out-of-domain benchmarks where the
best frozen system beats the fine-tuned baseline (ASV5 in EER and min-DCF; LA21
by paired bootstrap). \textbf{mDCF}: official ASVspoof~5 CM min-DCF.
WavLM-Large+MHFA (316M) is a matched-capacity probe (Section~\ref{sec:itw}),
outside the 2$\times$2.}
\label{tab:main}
\centering
\footnotesize
\renewcommand{\arraystretch}{1.2}
\setlength{\tabcolsep}{3pt}
\begin{tabular}{l c c c c c c c}
\toprule
\textbf{System} & \textbf{Tr.\ par.} & \textbf{LA19} & \textbf{LA21} &
\textbf{DF21} & \textbf{ITW} & \textbf{ASV5} & \textbf{mDCF}\\
\midrule
Fine-tuned (ft) & $\sim$316M
 & 2.54$\pm$1.71 & 11.75$\pm$0.91 & 6.34$\pm$3.62 & 24.37$\pm$6.37 & 27.81$\pm$2.15 & 0.77$\pm$0.05\\
\midrule
XEUS + MHFA & 0.39M
 & \textbf{2.12}$\pm$0.23 & \textbf{10.93}$\pm$1.26$^\dagger$
 & \textbf{7.97}$\pm$0.58 & \textbf{34.04}$\pm$2.98
 & \textbf{16.67}$\pm$0.78$^\dagger$ & \textbf{0.48}$\pm$0.02$^\dagger$\\
XEUS + AASIST & 0.33M
 & 2.23$\pm$0.43 & 14.61$\pm$2.43 & 9.16$\pm$2.57
 & 36.13$\pm$1.95 & 22.97$\pm$5.18 & 0.60$\pm$0.05\\
XLSR-53 + MHFA & 0.39M
 & 21.79$\pm$16.68 & 35.10$\pm$8.26 & 35.01$\pm$4.62
 & 40.46$\pm$7.77 & 39.55$\pm$7.97 & 0.90$\pm$0.10\\
XLSR-53 + AASIST & 0.33M
 & 15.08$\pm$4.66 & 26.09$\pm$0.73 & 29.03$\pm$3.02
 & 39.10$\pm$3.71 & 39.51$\pm$4.12 & 0.98$\pm$0.02\\
\midrule
WavLM-Large + MHFA & 0.39M
 & 4.47$\pm$0.47 & 22.43$\pm$3.74 & 9.73$\pm$0.75
 & 31.83$\pm$1.00 & 32.08$\pm$1.08 & 0.75$\pm$0.07\\
\bottomrule
\end{tabular}
\end{table*}

\subsection{Frozen encoder choice shapes cross-dataset behavior}
\label{sec:frozenbetter}
Table~\ref{tab:main} reports the frozen factorial as mean$\pm$std over three
seeds. With the encoder frozen, XEUS has the lower mean EER than XLSR-53 in
every cell. The difference is large and consistent on LA19, LA21, DF21, and
ASV5: holding the backend fixed to MHFA, frozen XEUS reduces error relative to
frozen XLSR-53 by roughly 58--77\% on LA21 (10.93 vs.\ 35.10), DF21 (7.97 vs.\
35.01), and ASV5 (16.67 vs.\ 39.55). On ITW the advantage is smaller (34.04 vs.\
40.46) and the seed distributions partly overlap, consistent with ITW being
among the hardest sets for every system. Within this recipe, the broader,
robustness-oriented pretraining of XEUS \cite{xeus} is associated with a fixed
representation that transfers, whereas frozen XLSR-53 largely does not. The
ordering also persists under matched aggregation: with pooling held fixed, XEUS
still has lower mean EER than XLSR-53 on every benchmark
(Section~\ref{sec:ablation}). We read this as an XEUS-specific instance, within
the present pipeline, of an emerging finding that the encoder, more than the
backend, governs cross-domain behavior \cite{serrano2025,spoofsuperb,raptor},
rather than as a new principle.

\textbf{Why is frozen XLSR-53 so variable in domain?} The frozen XLSR-53
in-domain EER is high and strongly seed-dependent (LA19 21.79$\pm$16.68 for
MHFA), and a reader may suspect a misconfigured cell. The variance is itself the
finding: across the three seeds, frozen XLSR-53+MHFA either converges to a
healthy in-domain floor (LA19 10.40, 14.03) or collapses toward chance (40.95),
with the same configuration, data, and frozen features differing only in
initialization. Two observations suggest that this instability is associated
with the frozen-XLSR setting rather than a general pipeline failure. First, the
same weights reach 0.85\% EER on LA19 once \emph{unfrozen} (best of three seeds;
three-seed mean 2.54 in Table~\ref{tab:main}), so the architecture, data, and
trainer are capable. Second, frozen XEUS passes through the identical fusion,
projections, and backend and trains stably (LA19 2.12$\pm$0.23). We therefore
report all systems as multi-seed mean$\pm$std rather than single draws; a
single-seed value for this recipe sits within this unstable distribution. Out
of domain, by contrast, frozen XLSR-53 is uniformly poor: even its best seed
remains far above frozen XEUS on LA21, DF21, and ASV5, so the encoder ordering
there is not an artifact of the in-domain instability.

\subsection{Encoder ordering persists under matched aggregation}
\label{sec:ablation}
The main factorial fixes each encoder in its best-practice aggregation
(Section~\ref{subsec:design}), so XEUS's advantage could in principle reflect
the encoder or merely the pooling choice. To isolate this, we retrain matched
MHFA cells in which, within each encoder, \emph{only} the aggregation differs
(final layer versus mean of the last four), holding everything else fixed over
the same three seeds (Table~\ref{tab:ablation}). In all ten matched
XEUS-versus-XLSR-53 comparisons (five datasets $\times$ two aggregations), XEUS
has the lower EER, both in three-seed mean and with non-overlapping 95\%
confidence intervals. The encoder effect therefore survives matched
aggregation.

\begin{table}[t]
\caption{Matched-aggregation ablation (MHFA backend, a separate three-seed run on
the same trainer; mean$\pm$std over seeds \{42, 1234, 2025\}). Within each
encoder, only the SSL layer-aggregation differs (final layer vs.\ mean of the
last four); all else is held fixed. \textbf{Bold} marks the lower EER (\%) in
each matched pair; XEUS is lower in all ten comparisons. This is a separate,
single-trainer study; absolute EERs are not commensurable with
Table~\ref{tab:main}, and the within-row XEUS-vs-XLSR ordering is the comparison
of interest.}
\label{tab:ablation}
\centering
\footnotesize
\renewcommand{\arraystretch}{1.15}
\setlength{\tabcolsep}{3pt}
\begin{tabular}{l cc cc}
\toprule
 & \multicolumn{2}{c}{\textbf{Final-layer}} & \multicolumn{2}{c}{\textbf{Mean-last-4}}\\
\cmidrule(lr){2-3}\cmidrule(lr){4-5}
\textbf{Dataset} & \textbf{XEUS} & \textbf{XLSR-53} & \textbf{XEUS} & \textbf{XLSR-53}\\
\midrule
LA19 & \textbf{1.96}$\pm$0.64  & 31.17$\pm$10.55 & \textbf{1.50}$\pm$0.24  & 27.61$\pm$15.83\\
LA21 & \textbf{12.16}$\pm$0.20 & 44.02$\pm$4.35  & \textbf{12.04}$\pm$0.51 & 37.79$\pm$8.68\\
DF21 & \textbf{8.24}$\pm$0.63  & 39.87$\pm$4.28  & \textbf{8.33}$\pm$0.65  & 35.89$\pm$3.86\\
ITW  & \textbf{27.87}$\pm$4.43 & 52.41$\pm$3.50  & \textbf{31.27}$\pm$0.79 & 44.46$\pm$3.12\\
ASV5 & \textbf{18.28}$\pm$1.45 & 44.81$\pm$3.82  & \textbf{23.46}$\pm$3.49 & 37.10$\pm$3.88\\
\bottomrule
\end{tabular}
\end{table}

Aggregation is a second-order effect whose sign is not consistent across
encoders: averaging the last four layers lowers XLSR-53's mean EER on all five
benchmarks, whereas for XEUS the effect is small and benchmark-dependent. It
helps in domain (LA19), falls within seed noise on LA21 and DF21, and favors the
final layer on the two far sets (ITW, ASV5). A factor that moves one encoder
uniformly but the other only inconsistently cannot produce the uniform
XEUS-over-XLSR ordering, which holds in all ten matched cells under both
aggregation schemes.

\subsection{Encoder and backend interact}
\label{sec:interact}
In domain, the two backends are comparable for frozen XEUS (MHFA 2.12 vs.\ AASIST
2.23, within seed noise) while AASIST is the more stable choice for frozen
XLSR-53 (15.08$\pm$4.66 vs.\ MHFA 21.79$\pm$16.68), so the in-domain backend
ranking already depends on the encoder. The interaction is sharper out of
domain. For frozen XEUS, MHFA is the better backend on every set (LA21 10.93 vs.\
14.61; DF21 7.97 vs.\ 9.16; ITW 34.04 vs.\ 36.13; ASV5 16.67 vs.\ 22.97), whereas
for frozen XLSR-53 AASIST matches or beats MHFA on every set (LA21 26.09 vs.\
35.10; DF21 29.03 vs.\ 35.01; ITW 39.10 vs.\ 40.46; ASV5 39.51 vs.\ 39.55).
Which backend works best therefore depends on the encoder it reads. A practical
reading is that MHFA exploits a strong fixed representation efficiently, while
AASIST's heavier relational modeling both salvages more signal from a weaker one
and stabilizes it: the high-variance cell is XLSR-53+MHFA (in-domain std 16.68),
and pairing the weak encoder with AASIST cuts that std to under a third (4.66).
That MHFA pairs well with a capable encoder is corroborated by the WeDefense
toolkit \cite{wedefense}, which reports MHFA over AASIST on PartialSpoof (0.80\%
vs.\ 1.04\%) and ASVspoof~5 (8.08\% vs.\ 17.11\%) with updated front ends, and by
MHFA's transfer across SSL models and tasks \cite{camhfa}. The implication is
that ``best backend'' is not encoder-agnostic, an interaction the baseline's
``different backend classifiers'' future-work item \cite{tvot} did not
anticipate, and consistent with large-scale evidence that composed
Pareto-optimal detector subsets beat naive ensembles \cite{stanek2026}.

\subsection{Frozen vs.\ full fine-tuning: stability and a partial inversion}
\label{sec:efficiency}
The in-pipeline fine-tuned baseline (XLSR-53 fully unfrozen + AASIST, three
seeds) lets us compare frozen detection against full fine-tuning under one
pipeline and scoring convention. In domain, the best fine-tuned seed reaches
0.85 on LA19, but the fine-tune is itself seed-unstable (2.54$\pm$1.71), so its
three-seed mean is comparable to frozen XEUS+MHFA's 2.12$\pm$0.23 rather than
clearly better. Out of domain the picture is mixed and dominated by fine-tune
seed variance. Frozen XEUS+MHFA beats the 316M-parameter fine-tune decisively on
ASVspoof~5 (16.67$\pm$0.78 vs.\ 27.81$\pm$2.15; non-overlapping intervals, an
11-point margin), an advantage that holds on the official ASV5 min-DCF
(0.48$\pm$0.02 vs.\ 0.77$\pm$0.05), and modestly on LA21 (10.93 vs.\ 11.75;
paired bootstrap, ft$-$frozen EER, $+0.82$ [$+0.64,+1.01$], excluding zero). On
the two mid-distance sets the fine-tune wins: DF21 by a paired bootstrap
($-1.63$ [$-1.88,-1.43$]) despite a far larger seed std (3.62 vs.\ 0.58), and
In-the-Wild more clearly (24.37$\pm$6.37 vs.\ 34.04$\pm$2.98; $-9.67$
[$-10.14,-9.24$]). Crucially, the fine-tune is markedly seed-unstable out of
domain: across three seeds its DF21 spans 4.04--10.52 and its ITW spans
18.93--31.37, so a single fine-tuned model's out-of-domain error is
unpredictable even where its mean wins. The frozen system, by contrast, has
lower observed seed variation on DF21, ITW, and ASV5 (LA21 is comparable, 1.26
vs.\ 0.91) and is ahead on the largest-scale benchmark while updating far fewer
parameters.

Within-run instability compounds this: probing the seed-42 run at epochs 5/10/15
(development EER 0.30/0.12/0.12; we report epoch 10, the earliest at the best
development EER), out-of-domain error does not track development EER, so
single-seed, single-checkpoint out-of-domain numbers in this regime are
unreliable. We did not reproduce the originating recipe's reported numbers
\cite{tvot}, obtained under a different pipeline. Our claim is the in-pipeline
comparison. Read as an exchange rate: a frozen representation read by a
sub-0.4M-parameter head updates roughly 800 times fewer parameters than a 316M
full fine-tune and is more seed-stable on most out-of-domain sets. It is ahead
on ASV5 and LA21, cedes DF21 and In-the-Wild, and is comparable in domain on
the three-seed mean. That freezing can rival fine-tuning out of domain has been
reported in fragments \cite{scalableaasist}; our contribution is the
in-pipeline, broad-coverage-encoder demonstration and the seed-stability
characterization.

\subsection{Scope: capacity and the choice of comparator}
\label{sec:itw}
Two caveats bound the encoder reading. First, scale, multilingual coverage, and
pretraining objective are confounded within our frozen pair: XEUS (577M, 4{,}057
languages, dereverberation objective \cite{xeus}) is larger and broader than
XLSR-53 (315M, 53 languages), so the matched-aggregation ablation shows that the
ordering persists across pooling choices but does not separate these three
axes. To probe whether XLSR-53's weak results are a generic consequence of using
a 315M frozen encoder, we add a frozen WavLM Large \cite{wavlm} + MHFA cell as a
capacity-matched diagnostic outside the $2\times2$ factorial (316M, matched to
XLSR-53's 315M, same pipeline). At matched capacity, WavLM Large outperforms
XLSR-53 out of domain by wide margins (LA21 22.43 vs.\ 35.10, DF21 9.73 vs.\
35.01, ASV5 32.08 vs.\ 39.55) and trains stably in domain (LA19 4.47$\pm$0.47
vs.\ 21.79$\pm$16.68). So 315M-scale frozen features do not inevitably produce
XLSR-53-level performance, and parameter count alone does not explain XLSR-53's
weaker performance relative to WavLM. This is consistent with concurrent
evidence that pretraining breadth may matter beyond raw parameter count
\cite{raptor}. WavLM Large does not, however, reach XEUS on most sets (XEUS wins
LA19, LA21, DF21, and ASV5). Since XEUS is also larger (577M vs.\ 316M), the
XEUS-over-WavLM margin stays confounded between scale and XEUS's pretraining,
which we do not claim to separate. WavLM Large does edge XEUS on ITW (31.83 vs.\
34.04), so broader coverage does not guarantee better performance in this
setting. ITW remains among the hardest settings for our frozen systems and the
out-of-domain set where the fine-tuned baseline's advantage is largest. Whether this reflects harder attacks or merely different ones
\cite{harderdiff} is open for frozen encoders, and found-in-the-wild robustness
remains a field-wide open problem.

\section{Discussion and Limitations}
\label{sec:limits}
\textbf{Pretraining/evaluation provenance.} Several benchmarks share provenance
with our encoders' pretraining corpora: LA19/LA21 and much of DF21 derive from
VCTK, and ASV5 is sourced from MLS, both of which appear in XEUS's pretraining.
We cannot fully separate representation-driven behavior from this exposure on
these sets, and we therefore scope our claim to the fixed downstream recipe:
changing the frozen encoder produces a large, repeatable change in cross-dataset
performance. Two observations inform, but do not eliminate, the concern. First,
XLSR-53 is itself MLS-pretrained (its pretraining is MLS-dominated) yet trails
XEUS on ASV5 by 23 EER points (16.67 vs.\ 39.55). Because both encoders were
exposed to MLS, corpus presence alone is insufficient to explain their ranking. However, this comparison does not quantify or exclude an effect of overlap.
Second, WavLM Large, a matched-capacity encoder that never saw VCTK, still clears
LA21 and DF21 by wide margins over XLSR-53 (22.43 vs.\ 35.10; 9.73 vs.\ 35.01),
so beating XLSR-53 on the VCTK-derived sets does not require VCTK exposure. The
advantage also holds on ASV5, which contains no VCTK. These observations weaken
a simple presence-or-absence explanation but do not estimate the contribution of
overlap to XEUS's performance. On the provenance-overlapping sets we accordingly
limit how strongly we speak of generalization, while the within-recipe
characterization stands. Isolating the residual contribution of exposure would
require provenance-clean benchmarks, which remain scarce.

\textbf{Encoder versus aggregation, scale, and objective.} Beyond encoder
identity the factorial also varies aggregation, epochs, and schedule. The aggregation factor is examined in Section~\ref{sec:ablation}, while the
matched-capacity WavLM Large comparison in Section~\ref{sec:itw} partially
probes scale, leaving multilingual coverage and the dereverberation objective
entangled---a full sweep of which is future work.

\textbf{Pipeline, seeds, and augmentation.} All systems are trained and scored in
one pipeline over three seeds. Frozen XLSR-53 is seed-unstable in domain and the
fine-tune out of domain, so multi-seed reporting is necessary rather than
cosmetic for both. Following the baseline, we use no augmentation and train only on LA19. RawBoost-style augmentation \cite{rawboost} and broader data would
likely help ITW and ASV5 for frozen and fine-tuned systems alike. This shared
regime, not under-training, explains why our XLSR-53 systems sit above some
externally reported levels, and our claims are internal to this one pipeline.

\textbf{Prosody.} The frozen-versus-frozen results are consistent with the
prosody motivation, but the design cannot attribute the gain to prosody
specifically rather than to general pretraining breadth, so we frame it
as suggestive rather than as a demonstrated mechanism.

\section{Conclusion}
Freezing the SSL encoder turns deepfake detection into a characterization
problem: where does cross-domain behavior come from? Within this recipe, much of
it tracks the frozen encoder. Swapping it lowers cross-dataset error by more than
half on LA21, DF21, and ASV5 under the MHFA backend, and the ordering persists
in a matched-aggregation ablation. A capacity-matched WavLM Large diagnostic
shows that XLSR-53's weaker performance cannot be explained by parameter count alone. Our design cannot separate broader coverage from the dereverberation objective, and because several benchmarks
share provenance with encoder pretraining we scope the result to the recipe
rather than claim distributional generalization. Some of it lies in the pairing:
backends do not rank independently of the encoder they read, so ``best
backend'' is not a transferable fact. And fine-tuning is not a uniform win:
against a three-seed 316M full fine-tune in the same pipeline, a frozen
broad-coverage encoder read by a sub-0.4M-parameter head is comparable in
domain, has lower observed seed variation on most out-of-domain sets, and wins
on LA21 and ASVspoof~5, while the fine-tune takes DF21 and In-the-Wild.

These are also deployment questions: detectors fielded in privacy-sensitive
settings may not be re-fine-tuned per condition, so trustworthy deployment
requires knowing which frozen representations transfer and how to read them
efficiently. Characterization is the groundwork for that, not a retreat from
performance.

\section*{Acknowledgments}
This work was supported in part by computational resources provided by the
NSF ACCESS program under allocation CIS261055. We thank James O'Brien of
Sound Ethics for his mentorship.

OpenAI ChatGPT and Anthropic Claude were used for language editing and
rephrasing in the Abstract, Introduction, Related Work, Results and Discussion,
and Conclusion. The authors reviewed and edited all AI-assisted text and take
full responsibility for the final content.


\begin{thebibliography}{00}

\bibitem{tak}
H.~Tak, M.~Todisco, X.~Wang, J.-w. Jung, J.~Yamagishi, and N.~Evans, ``Automatic
speaker verification spoofing and deepfake detection using wav2vec 2.0 and data
augmentation,'' in \emph{Proc. Odyssey}, 2022, pp.~112--119.

\bibitem{wangyama}
X.~Wang and J.~Yamagishi, ``Investigating self-supervised front ends for speech
spoofing countermeasures,'' in \emph{Proc. Odyssey}, 2022, pp.~100--106.

\bibitem{liu2023asvspoof21}
X.~Liu, X.~Wang, M.~Sahidullah, J.~Patino, H.~Delgado, T.~Kinnunen,
M.~Todisco, J.~Yamagishi, N.~Evans, A.~Nautsch, and K.~A. Lee, ``ASVspoof 2021:
towards spoofed and deepfake speech detection in the wild,'' \emph{IEEE/ACM
Trans. Audio, Speech, Language Process.}, vol.~31, pp.~2507--2522, 2023.

\bibitem{muller2022}
N.~M.~M\"uller, P.~Czempin, F.~Dieckmann, A.~Froghyar, and K.~B\"ottinger,
``Does audio deepfake detection generalize?,'' in \emph{Proc. Interspeech}, 2022,
pp.~2783--2787.

\bibitem{tvot}
Y.~El~Kheir, A.~Das, E.~E. Erdogan, F.~Ritter-Guti\'errez, T.~Polzehl, and
S.~M\"oller, ``Two views, one truth: spectral and self-supervised features
fusion for robust speech deepfake detection,'' in \emph{Proc. IEEE Workshop
Appl. Signal Process. Audio Acoust. (WASPAA)}, Tahoe City, CA, USA, 2025,
pp.~1--5, doi: 10.1109/WASPAA66052.2025.11230938.

\bibitem{xeus}
W.~Chen, W.~Zhang, Y.~Peng, X.~Li, J.~Tian, J.~Shi, X.~Chang, S.~Maiti,
K.~Livescu, and S.~Watanabe, ``Towards robust speech representation learning
for thousands of languages,'' in \emph{Proc. EMNLP}, 2024, pp.~10205--10224.

\bibitem{xlsr}
A.~Conneau, A.~Baevski, R.~Collobert, A.~Mohamed, and M.~Auli, ``Unsupervised
cross-lingual representation learning for speech recognition,'' in \emph{Proc.
Interspeech}, 2021, pp.~2426--2430.

\bibitem{doubleentendre}
M.~Frohmann, G.~Meseguer-Brocal, M.~Schedl, and E.~V. Epure, ``Double
entendre: robust audio-based AI-generated lyrics detection via multi-view
fusion,'' in \emph{Findings of ACL}, 2025, pp.~1914--1926.

\bibitem{serrano2025}
P.~Serrano, R.~Duroselle, F.~Angulo, J.-F. Bonastre, and O.~Boeffard,
``Improving out-of-domain audio deepfake detection via layer selection and
fusion of SSL-based countermeasures,'' arXiv:2509.12003, 2025.

\bibitem{afchar2024}
D.~Afchar, G.~Meseguer-Brocal, and R.~Hennequin, ``Detecting music deepfakes is
easy but actually hard,'' arXiv:2405.04181, 2024.

\bibitem{wedefense}
L.~Zhang, J.~Rohdin, X.~Wang, J.~Peng, T.~Liu, Y.~Zhang, H.-T. Luong, S.~Wang,
C.~Liang, A.~Silnova, and N.~Evans, ``WeDefense: a toolkit to defend against
fake audio,'' arXiv:2601.15240, 2026.

\bibitem{hu2022lora}
E.~J. Hu, Y.~Shen, P.~Wallis, Z.~Allen-Zhu, Y.~Li, S.~Wang, L.~Wang, and
W.~Chen, ``LoRA: low-rank adaptation of large language models,'' in \emph{Proc.
ICLR}, 2022.

\bibitem{aasist3}
K.~Borodin, V.~Kudryavtsev, D.~Korzh, A.~Efimenko, G.~Mkrtchian,
M.~Gorodnichev, and O.~Y.~Rogov, ``AASIST3: KAN-enhanced AASIST speech
deepfake detection using SSL features and additional regularization for the
ASVspoof 2024 challenge,'' in \emph{Proc. ASVspoof 2024 Workshop (Autom.
Speaker Verif. Spoofing Countermeas.)}, Kos, Greece, 2024, pp.~48--55,
doi: 10.21437/ASVspoof.2024-8.

\bibitem{spoofsuperb}
H.~Ali, N.~S. Adupa, S.~Subramani, and H.~Malik, ``A SUPERB-style benchmark of
self-supervised speech models for audio deepfake detection,'' in \emph{Proc.
IEEE Int. Conf. Acoust., Speech Signal Process. (ICASSP)}, 2026,
arXiv:2603.01482.

\bibitem{aasist}
J.-w. Jung, H.-S. Heo, H.~Tak, H.-j. Shim, J.~S. Chung, B.-J. Lee, H.-J. Yu,
and N.~Evans, ``AASIST: audio anti-spoofing using integrated spectro-temporal
graph attention networks,'' in \emph{Proc. IEEE ICASSP}, 2022, pp.~6367--6371.

\bibitem{mhfa}
J.~Peng, O.~Plchot, T.~Stafylakis, L.~Mo\v{s}ner, L.~Burget, and
J.~\v{C}ernock\'y, ``An attention-based backend allowing efficient fine-tuning
of transformer models for speaker verification,'' in \emph{Proc. IEEE SLT},
2022, pp.~555--562.

\bibitem{pasad2021}
A.~Pasad, J.-C. Chou, and K.~Livescu, ``Layer-wise analysis of a
self-supervised speech representation model,'' in \emph{Proc. IEEE ASRU}, 2021,
pp.~914--921.

\bibitem{elkheir2025layer}
Y.~El~Kheir, Y.~Samih, S.~Maharjan, T.~Polzehl, and S.~M\"oller,
``Comprehensive layer-wise analysis of SSL models for audio deepfake
detection,'' in \emph{Findings of NAACL}, 2025, pp.~4070--4082.

\bibitem{nes2net}
T.~Liu, D.-T. Truong, R.~K. Das, K.~A. Lee, and H.~Li, ``Nes2Net: a lightweight
nested architecture for foundation model driven speech anti-spoofing,''
\emph{IEEE Trans. Inf. Forensics Security}, vol.~20, pp.~12005--12018, 2025,
doi: 10.1109/TIFS.2025.3626963.

\bibitem{dao2026}
A.-T. Dao, D.~Matrouf, M.~Rouvier, and N.~Evans, ``Enhancing multi-corpus
training in SSL-based anti-spoofing models: domain-invariant feature
extraction,'' arXiv:2603.18657, 2026.

\bibitem{thani2026}
U.~Thani, G.~Singh, and P.~Singh, ``Unsupervised domain adaptation for audio
deepfake detection with modular statistical transformations,'' in \emph{Proc.
IEEE/CVF Conf. Comput. Vis. Pattern Recognit. Workshops (CVPRW)}, 2026,
pp.~11031--11039.

\bibitem{li2026mlitw}
D.~Li, J.~Xue, Y.~Ren, Z.~Yi, Y.~Huang, G.~Feng, and Y.~Chai, ``How well do
current speech deepfake detection methods generalize to the real world?,''
arXiv:2603.05852, 2026.

\bibitem{harderdiff}
N.~M.~M\"uller, N.~Evans, H.~Tak, P.~Sperl, and K.~B\"ottinger, ``Harder or
different? Understanding generalization of audio deepfake detection,'' in
\emph{Proc. Interspeech}, Kos, Greece, 2024, pp.~2705--2709,
doi: 10.21437/Interspeech.2024-247.

\bibitem{cqcc}
M.~Todisco, H.~Delgado, and N.~Evans, ``Constant Q cepstral coefficients: a
spoofing countermeasure for automatic speaker verification,'' \emph{Comput.
Speech Lang.}, vol.~45, pp.~516--535, 2017.

\bibitem{asvspoof2019}
X.~Wang \emph{et~al.}, ``ASVspoof 2019: a large-scale public database of
synthesized, converted and replayed speech,'' \emph{Comput. Speech Lang.},
vol.~64, 2020, Art. no.~101114, doi: 10.1016/j.csl.2020.101114.

\bibitem{wang2024asvspoof5}
X.~Wang, H.~Delgado, H.~Tak, J.-w. Jung, H.-j. Shim, M.~Todisco, I.~Kukanov,
X.~Liu, M.~Sahidullah, T.~H. Kinnunen, N.~Evans, K.~A. Lee, and J.~Yamagishi,
``ASVspoof 5: crowdsourced speech data, deepfakes, and adversarial attacks at
scale,'' in \emph{Proc. ASVspoof 2024 Workshop (Autom. Speaker Verif. Spoofing
Countermeas.)}, Kos, Greece, 2024, pp.~1--8, doi: 10.21437/ASVspoof.2024-1.

\bibitem{raptor}
A.~Kulkarni, S.~Dowerah, A.~Kulkarni, T.~Alum\"ae, and M.~Magimai~Doss,
``Do compact SSL backbones matter for audio deepfake detection? A controlled
study with RAPTOR,'' 2026, arXiv:2603.06164.

\bibitem{camhfa}
J.~Peng, L.~Mo\v{s}ner, L.~Zhang, O.~Plchot, T.~Stafylakis, L.~Burget, and
J.~\v{C}ernock\'y, ``CA-MHFA: a context-aware multi-head factorized attentive
pooling for SSL-based speaker verification,'' in \emph{Proc. IEEE ICASSP},
2025, pp.~1--5.

\bibitem{stanek2026}
V.~Stan\v{e}k, M.~Pere\v{s}\'ini, L.~Sekanina, A.~Firc, and K.~Malinka,
``Evolutionary multi-objective fusion of deepfake speech detectors,'' in
\emph{Proc. IEEE Congr. Evol. Comput. (CEC)}, 2026, arXiv:2604.01330.

\bibitem{scalableaasist}
I.~Viakhirev, D.~Sirota, A.~Smirnov, and K.~Borodin, ``Towards scalable
AASIST: refining graph attention for speech deepfake detection,'' 2025,
arXiv:2507.11777.

\bibitem{wavlm}
S.~Chen \emph{et~al.}, ``WavLM: large-scale self-supervised pre-training for
full stack speech processing,'' \emph{IEEE J. Sel. Topics Signal Process.},
vol.~16, no.~6, pp.~1505--1518, Oct. 2022.

\bibitem{rawboost}
H.~Tak, M.~R. Kamble, J.~Patino, M.~Todisco, and N.~Evans, ``RawBoost: a raw
data boosting and augmentation method applied to automatic speaker verification
anti-spoofing,'' in \emph{Proc. IEEE ICASSP}, 2022, pp.~6382--6386.

\end{thebibliography}
\end{document}